\documentclass[prd,aps,twocolumn,superscriptaddress,preprintnumbers,showpacs,tightenlines,floatfix,amssymb]{revtex4}
\usepackage{epsfig}

\newcommand{\beqn}{\begin{eqnarray}}
\newcommand{\eeqn}{\end{eqnarray}}
\newcommand{\eq}[1]{(\ref{#1})}
\newcommand{\cC}{{\cal C}}
\newcommand{\cZ}{{\cal Z}}
\newcommand{\dd}{{\mathrm d}}
\newcommand{\dD}{{\mathrm D}}
\newcommand{\Tr}{{\mathrm{Tr}}\,}

\def\bbbone{{\mathchoice {\rm 1\mskip-4mu l} {\rm 1\mskip-4mu l}
{\rm 1\mskip-4.5mu l} {\rm 1\mskip-5mu l}}}

\begin{document}

\title{Thermal monopoles and selfdual dyons in the Quark-Gluon Plasma}

\author{M.N.~Chernodub}
\affiliation{
CNRS, LMPT,
F\'ed\'eration Denis Poisson, Universit\'e de Tours,
F-37200, Tours, France}
\affiliation{
DMPA,
University of Gent, Krijgslaan 281, S9, B-9000 Gent, Belgium}
\affiliation{
ITEP,
B.~Cheremushkinskaya 25, Moscow, 117218, Russia}
\author{A.~D'Alessandro}
\affiliation{Dipartimento di Fisica, Universit\`a di Genova and INFN,
Via Dodecaneso 33, I-16146 Genova, Italy}
\author{M.~D'Elia}
\affiliation{Dipartimento di Fisica, Universit\`a di Genova and INFN,
Via Dodecaneso 33, I-16146 Genova, Italy}
\author{V.I.~Zakharov}
\affiliation{
ITEP,
B.~Cheremushkinskaya 25, Moscow, 117218, Russia}
\affiliation{Max-Planck-Institut f\"ur Physik, F\"ohringer Ring 6, 80805 M\"unich, Germany}

\preprint{ITEP-LAT/2009-11}
\preprint{GEF-TH 10/09}

\begin{abstract}
We perform a numerical study of the excess of non-abelian gauge invariant gluonic action around thermal abelian monopoles
which populate the deconfined phase of Yang-Mills theories. Our results show that the excess of magnetic action is
close to that of the electric one,
so that thermal abelian monopoles may be associated with physical objects carrying both
electric and magnetic charge, i.e. dyons. Thus, the quark gluon plasma is likely to be populated by
selfdual dyons, which may manifest themselves in the heavy-ion collisions via the chiral magnetic effect.
Thermodynamically, thermal monopoles provide a negative contribution to the pressure of the system.
\end{abstract}


\date{September 26, 2009}

\maketitle

\section{Introduction}

Interpretation of heavy ion experiments at RHIC and first-principle lattice
simulations suggest that the quark-gluon plasma has quite unusual
properties~\cite{review}. Contrary to general expectations, at temperatures
just above the critical temperature, $T_c$, the plasma looks like an
ideal fluid rather than a system of weakly interacting quark-gluon gas.
The reason of this behavior lies in the strongly interacting  nature of the
quark-gluon plasma.

The magnetic component of the gluon fields may be a key player in
the unusual dynamics of the plasma~\cite{ref:magn:comp,shuryak,ref:evidence,EOS:vortices,EOS:magnetic,sasha,ref:magn:strings,ref:viscosity}.
The basic idea underlying the magnetic picture can be motivated in the following way~\cite{ref:magn:comp}.
As it was suggested long time ago in Ref.~\cite{ref:dual}, at low temperatures the
magnetic degrees of freedom exist in the form of a (gluo)magnetic condensate.
Magnetic charge condensation is believed to be responsible for color confinement and this is usually known as the dual superconductor
scenario: the existence of such condensation and its disappearance at the critical temperature $T_c$
has been verified in various lattice QCD studies (see e.g. Refs.~\cite{superI-II,superIV,moscow,bari}
or Ref.~\cite{greensite} for a review). As the temperature increases, the condensate
starts to evaporate and it is destroyed
completely at $T_c$, while the magnetic color degrees are released as thermal states into the gluon plasma.
The central point is that simultaneously the magnetic degrees of freedom are to become
physical and constitute a real, thermally excited component of the gluon plasma~\cite{ref:magn:comp}.

The magnetic monopoles in the Yang-Mills plasma are particlelike objects made of gluons.
The monopoles are suggested to affect both the transport properties and the thermodynamical
features of the plasma~\cite{ref:magn:comp,shuryak,ref:viscosity}. This suggestion can in principle
be checked by numerical simulations of Yang-Mills theory on the lattice. The
lattice method provides us with thermalized ensembles of monopole trajectories
in Euclidean space time. However, in order to calculate the contribution of thermal
(i.e., real) monopoles to any physical quantity one should be able to separate the
thermal monopoles from numerous virtual monopole loops. Virtual monopoles -- which are
responsible essentially for ultraviolet physics -- are known to populate the thermal
ensembles in numerous quantities, so that the separation of monopoles into real and
virtual parts seems to be a difficult task.

It was shown in Ref.~\cite{ref:magn:comp} that thermal monopoles in Minkowski
space-time are associated with Euclidean monopole trajectories that have a nontrivial
wrapping number on the short (thermal) direction of the Euclidean space.
Moreover, the density of the monopoles in the Minkowski space-time is given by
the thermal average of the absolute value of the monopole wrapping number.
The properties of the wrapped trajectories were studied numerically using lattice simulations
in Ref.~\cite{ref:preliminary,ref:evidence}. In Ref.~\cite{ref:evidence} the
average wrapping number of the monopole trajectories was thoroughly  calculated. It was shown
that this quantity is free from ultraviolet artifacts as it should be the case
for a purely thermal quantity. The monopole density grows with the temperature $T$.
However, the parametrical $T$-dependence of this quantity differs from the behavior
expected in the case of an ideal gas of the monopoles. This is one of the first lattice
evidences on the nontrivial dynamics of the magnetic monopoles in the high temperature
gauge theory~\cite{ref:evidence}.

Monopoles do not appear alone in the plasma. Numerical simulations have revealed the fact
that monopoles are a part of chainlike structures~\cite{greensite}. In these structures the
(anti)monopoles are connected to each other by magnetic vortices. The existence of the
vortex defects is related to the nontriviality of the center of the gauge group. Both the
monopoles and vortices contribute to the equation of state of Yang-Mills theory
at finite temperature~\cite{EOS:magnetic,EOS:vortices} (a short review of the subject
can be found in Ref.~\cite{ref:magn:strings}).

A calculation of the effect of the vortices on the pressure of the Yang-Mills plasma
can be formulated in a rather straightforward and clear way. According to Ref.~\cite{EOS:vortices}
the vortices provide a negative contribution to the pressure in agreement with theoretical
expectations of Ref.~\cite{sasha}. As for the monopoles, the preliminary calculation of Ref.~\cite{EOS:magnetic}
indicated that they provide a large positive contribution. However, the calculation was done using a nonlocal
method which did not discriminate between the real and virtual monopole loops. Below we show
that the contribution of the {\it thermal} monopoles to the pressure of gluons is negative as well.

Thermal monopoles detected by lattice QCD simulations are of abelian nature. An abelian projection
is needed to define magnetic monopole currents (see the Appendix for details) and the Maximal Abelian Gauge
(MAG) is usually chosen to do that. The fact that both the density and the spatial correlations
of thermal monopoles measured in the MAG projection are free of UV artifacts~\cite{ref:evidence} (i.e. they
correctly scale to the continuum limit) points to an underlying physical meaning of those objects.
Nevertheless, the dependence on the chosen abelian projection still poses a problem.
In order to further investigate this issue and better clarify the physical properties of thermal monopoles,
we have decided to measure the correlation of the unprojected gauge invariant nonabelian action density
with the positions of abelian thermal monopoles. As we shall clarify in the following, this is also the essential
quantity entering the determination of thermodynamic properties of thermal monopoles (e.g. their contribution
to the pressure).

We made refined measurements which treat the magnetic and electric
parts of the gluon action separately. This allowed us to find -- as
an unexpected byproduct -- that the excesses of the magnetic and
electric action densities have almost the same value near the
monopoles suggesting that the electric and magnetic components of
these objects have the same significance. We conclude that the
objects which we identified simply as thermal abelian monopoles
after abelian projection, may correspond to more interesting
physical objects, actually thermal dyons, with equal electric and
magnetic charges~\cite{footnote:dyon}

The dyons in zero-temperature Euclidean Yang-Mills theory were discussed in the literature for a long time both
in continuum~\cite{ref:dyons:coninuum} and lattice~\cite{ref:dyons:lattice,ref:dyonic:lattice} formulations.
In Ref.~\cite{ref:dyonic:lattice} smooth dyonic solutions were observed in the deconfinement phase of lattice
Yang-Mills theory. Here we discuss another lattice evidence in favor of existence of dyons as real objects in
quark gluon plasma. It is necessary to mention that these dyons would have a finite density which
is independent of the ultraviolet cutoff.

The importance of non-Abelian dyons in the quark gluon plasma is difficult to overestimate since these objects may play
a key role in the so-called chiral magnetic effect. { This effect -- which reveals certain $CP$-odd correlations
in the quark-gluon plasma -- was suggested theoretically in Ref.~\cite{Kharzeev:08:1}. The chiral magnetic effect
leads to generation of an electric current along the direction of a
magnetic field in a nontrivial topological backgrounds of gluons.} The strong magnetic field is
naturally created in heavy-ion collisions, and the appearance of the electric current is detected in a form of a certain
geometric asymmetry  of electrically charged particles produced in the collisions.
There are preliminary indications that the chiral magnetic effect has been
indeed observed by the STAR Collaboration in experiments at RHIC \cite{Voloshin:08:1}. The signatures of this
phenomenon were also found in recent numerical lattice simulations~\cite{ref:CME:Lattice}, and discussed
in nonperturbative holographic approaches of Ref.~\cite{Yee:2009vw}. In case the dyons are
really associated to the thermal monopoles, as suggested by our results, they
may serve as fluctuations of the topological charge in the quark gluon plasma phase.

The structure of this paper is as follows. In Sec.~\ref{sec:thermo} we review basic thermodynamical
relations of Yang-Mills theory. In Sec.~\ref{sec:monopoles} we formulate how to estimate the influence
of the monopoles on thermodynamics and how to calculate the electric and magnetic action densities around the monopoles.
In Sec.~\ref{sec:lattice} we present our numerical results. Our conclusion is summarized in the last Section,
{ and technical details are given in the Appendix.}

\section{Thermodynamics of gluons}
\label{sec:thermo}

The partition function of the Yang--Mills theory is
\beqn
\cZ = \int \dD A \, \exp\left\{- \frac{1}{2 g^2} \Tr \, G_{\mu\nu}^2 \right\}\,,
\label{eq:cZ}
\eeqn
where $G_{\mu\nu} = G_{\mu\nu}^a t^a$ is the field strength tensor of the non-Abelian field $A$, and
$t^a$ are the $SU(N)$ generators normalized in the standard way, $\Tr t^a t^b = \frac{1}{2} \delta^{ab}$.

For a sufficiently large and homogeneous gluon system
residing in thermodynamical equilibrium the pressure $p$ and the energy density $\varepsilon$
are given by the derivatives of the partition
function~\eq{eq:cZ} with respect to the spatial volume, $V$,
and the temperature~$T$, respectively,
\beqn
p & = & \frac{T}{V} \frac{\partial \log Z(T,V)}{\partial \log V} = \frac{T}{V} \log \cZ(T,V)\,,
\label{eq:pressure}
\\
\varepsilon & = & \frac{T}{V} \frac{\partial \log Z(T,V)}{\partial \log T}\,,
\label{eq:energy}
\eeqn

Following a standard approach (see, e.g., Ref.~\cite{Boyd:1996bx}), one can relate
the energy density $\varepsilon$ and the pressure $p$ can via the quantum
average the trace of the energy--momentum tensor $T_{\mu\nu}$,
\beqn
\theta(T) = \langle T^\mu_\mu \rangle \equiv \varepsilon - 3 p\,,
\label{eq:theta}
\eeqn
where in Yang--Mills theory
\beqn
T_{\mu\nu} = 2 \Tr \left[G_{\mu\sigma} G_{\nu\sigma} - \frac{1}{4} \delta_{\mu\nu} G_{\sigma\rho} G_{\sigma\rho}\right]\,.
\label{eq:T}
\eeqn
At the classical level trace of this quantity is zero because the {\it bare} Yang--Mills theory is a conformal theory.
However, at the quantum level the conformal invariance is broken, and, consequently, the energy--momentum tensor exhibits
a trace anomaly: the quantum average of the trace of the energy momentum tensor~\eq{eq:theta} is nonzero.
The trace anomaly is intimately related to the gluon condensate which breaks the scale invariance of the theory
\beqn
\theta(T) = \Bigl\langle \frac{\tilde\beta(g)}{2 g} G_{\mu\nu}^a G_{\mu\nu}^a \Bigr\rangle\,,
\label{eq:theta:continuum}
\eeqn
where $\tilde\beta(g)$ the Gell-Mann-Low $\beta$-function. We use the notation $\tilde\beta$ instead
of the conventional $\beta$ in order to avoid a confusion with the lattice coupling constant
\beqn
\beta = \frac{2 N}{g^2}\,.
\label{eq:beta}
\eeqn

The trace anomaly~\eq{eq:theta} is an important thermodynamic quantity because it allows us to reconstruct
both  the pressure and the energy density as follows:
\beqn
p(T) & = & T^4 \int\nolimits^T_0 \ \frac{\dd\, T_1}{T_1} \ \frac{\theta(T_1)}{T_1^4}\,,
\label{eq:pressure:anomaly}\\
\varepsilon(T) & = & 3 \, T^4 \int\nolimits^T_0 \ \frac{\dd\, T_1}{T_1} \ \frac{\theta(T_1)}{T_1^4} + \theta(T)\,.
\label{eq:energy:anomaly}
\eeqn

On the lattice the dynamical fields are the gluonic $SU(N)$ matrices $U_{x\mu}$ which
are identified at the lattice links $l=\{x,\mu\}$. The link gauge fields are related
to the continuum gauge fields $A_\mu(a)$ in the limit of vanishing lattice spacing, $a \to 0$:
\beqn
U_{x\mu} & = & \exp\Bigl\{i g \int\limits_x^{x+a \hat \mu} \dd y \, A_\mu(y) \Bigr\} \\
& \to & \bbbone + i a g A_\mu(x) + O(a^2)\,,
\nonumber
\eeqn
The lattice spacing $a$ is a function of the lattice coupling~\eq{eq:beta}.

The lattice analogue of the partition function~\eq{eq:cZ} is
\beqn
\cZ(T,V) = \int D U \, \exp\Bigl\{ - \beta \sum_P S_P[U]\Bigr\}\,.
\label{eq:lattice:cZ}
\eeqn
where the plaquette action density $S_P[U]$ is usually written in the Wilson form,
\beqn
S_P[U] = 1 - \frac{1}{N} {\mathrm{Re}}\, \Tr U_P\,, \qquad U_P = \prod_{l \in \partial P} U_l\,,
\label{eq:SP}
\eeqn

The spatial volume $V = (L_s a)^3$ and the temperature
\beqn
T = \frac{1}{L_t a}
\label{eq:temperature}
\eeqn
of the gluonic system are defined by the asymmetric geometry of
the Euclidean lattice, $L_s^3 L_t$. The shorter direction, $L_t$ with $L_t \leqslant L_s$, is associated with the imaginary time,
or, ``temperature'' direction.  The trace anomaly~\eq{eq:theta:continuum} on the lattice is written as follows
\beqn
\frac{\theta(T)}{T^4} = 6 \, L_t^4 \left(\frac{\partial \, \beta(a)}{\partial \log a} \right)
\cdot \Bigl(\langle S_P \rangle_T - \langle S_P \rangle_0\Bigr)\,.
\label{eq:anomaly:lattice}
\eeqn
where ${\langle S_P \rangle}_T$ and ${\langle S_P \rangle}_0$ are the action densities at $T>0$ and at $T=0$.
The subtraction of the $T=0$ value is needed to
remove the effect of quantum fluctuations, which lead to ultraviolet (UV) divergency of the quantum expectation
value. The regularized trace anomaly~\eq{eq:anomaly:lattice} is an UV-finite quantity.

We used the Monte Carlo simulations on the lattice to evaluate the influence of the monopoles on the trace anomaly,
and, consequently, on the equation of state in the thermal Yang-Mills theory. We describe setup of our
numerical simulations in Appendix~\ref{sec:numerical}.

\section{Local action densities}
\label{sec:monopoles}

The (chromo)magnetic monopoles in the Yang-Mills plasma are particlelike objects made of gluons which
are generally believed to be responsible for color confinement in QCD. The monopole
confinement scenario is reviewed in Ref.~\cite{greensite}. On the lattice the monopoles
appear in the form of the closed trajectories. The monopole loops may be either wrapped around
the short (imaginary time) direction of the lattice, or they may form shrinkable loops with
zero winding number. The monopole configurations with nontrivial winding are associated with
thermal  monopoles~\cite{ref:magn:comp}, while the other trajectories correspond to virtual loops.
The properties of the winding trajectories were studied long ago in Ref.~\cite{ref:bornyakov,ref:ejiri}.
A recent study~\cite{ref:evidence} has revealed that the thermal monopole density,
\beqn
\rho = \frac{1}{V_{3D}}
\left\langle \sum_{\vec{x}} \left| N_{\mathrm{wrap}}[m_0(\vec{x},t)]\right| \right\rangle \,,
\quad V_{3D} = L_s^3\,,
\label{densdef}
\eeqn
identified as a density of the average wrapping number~\cite{ref:magn:comp} $N_{\mathrm{wrap}}[m_0(\vec{x},t)]$,
shows a very { good} scaling towards a continuum limit. This result confirms the physical nature of thermal,
or real, magnetic monopoles in the Yang-Mills plasma.

Expression~\eq{eq:theta:continuum} implies that if the monopoles affect the gluonic condensate
then the monopoles should also make influence on the trace anomaly and, as a consequence, on the
equation of state of the gluon plasma. Translated to the lattice language~\eq{eq:anomaly:lattice},
this statement means that the thermal monopoles contribute to equation of state if they affect the
expectation value of the plaquette action~$S_P$.

At zero temperature the correlation of the monopoles with the action density was indeed established
in Refs.~\cite{ref:ITEP:M,ref:ITEP:E,ref:anatomy}. Although this result was obtained in a cold vacuum,
it provides us with a hint to suspect that the monopoles may contribute to the equation of state at
finite temperature as well.

In the continuum limit, the averaged value of the action density at the position of monopoles can
be calculated with the help of the normalized correlator between the monopole trajectory $j_\mu(x)$
and the non-Abelian action density,
\beqn
\langle S \rangle_{\mathrm{mon}}(x,x') & \equiv & \frac{1}{2 g^2} \langle \Tr G_{\mu\nu}^2 \rangle_{\mathrm{mon}}(x,x')
\label{eq:total}\\
& = & \frac{1}{\langle j^2_\beta(x)\rangle} \,
\Bigl\langle j^2_\alpha(x) \frac{1}{2 g^2} \Tr G_{\mu\nu}^2(x') \Bigr\rangle\,.
\nonumber
\eeqn
The points $x$ and $x'$ are very close to each other. As we will see below the lattice provides a natural
regularization of the correlation function in Eq.~\eq{eq:total}.

The contribution of the monopoles~\eq{eq:total} to the gluon energy density (or, equivalently, to the gluonic condensate)
can be subdivided into the electric and magnetic parts, respectively:
\beqn
\langle S \rangle_{\mathrm{mon}} = \langle S_M \rangle_{\mathrm{mon}} + \langle S_E \rangle_{\mathrm{mon}}\,.
\eeqn
The magnetic part of the action density can naturally be expressed as follows~\cite{ref:ITEP:M}:
\beqn
\langle S_M\rangle_{\mathrm{mon}}(x,x') = \frac{1}{\langle j^2_\alpha(x)\rangle} \,
\left\langle \frac{1}{2 g^2} \Tr {[j_\mu(x) \, {\widetilde G}_{\mu\nu}(x')]}^2 \right\rangle\
\label{eq:magnetic}
\eeqn
where
\beqn
{\widetilde G}_{\mu\nu} (x) = \frac{1}{2} \varepsilon_{\mu\nu\alpha\beta} G_{\alpha\beta} (x)\,.
\eeqn
If the monopole is static, $j_\mu \sim \delta_{\mu 4}$ then only spacelike (magnetic) components of the
field strength tensor $G_{ij}$ with $i,j=1,2,3$ contribute to the correlator~\eq{eq:magnetic}.

The excess of the chromoelectric action around the monopoles with respect to the chromoelectric vacuum expectation
value in the bulk can be evaluated with the help of the following normalized correlator~\cite{ref:ITEP:E}:
\beqn
\langle S_E\rangle_{\mathrm{mon}}(x,x') = \frac{1}{\langle j^2_\alpha(x)\rangle} \,
\left\langle \frac{1}{2} \Tr {[j_\mu(x) \, G_{\mu\nu}(x')]}^2 \right\rangle\,.
\label{eq:electric}
\eeqn
If the monopole is static, $j_\mu \sim \delta_{\mu 4}$ then only timelike (electric) components of the
field strength tensor $G_{4i}$ with $i=1,2,3$ contribute to the correlator~\eq{eq:electric}.

\section{Lattice results}
\label{sec:lattice}

\subsection{Lattice observables}

The continuum definitions \eq{eq:magnetic} and \eq{eq:electric} can be translated to the lattice as follows.
First, we locate the wrapped monopole trajectories using a standard prescription described in Appendix~\ref{sec:numerical}.
The monopole current is defined on the links $\{x,\mu\}^*$ of the dual lattice and takes integer values, $j_\mu(x) = 0, \pm 1, \dots$.
Usually the multiple-charged currents with $|j_\mu(x)| \geqslant 2$ are very rare, so that one can set $ |j_\mu(x)| = j^2_\mu(x)$
with a high accuracy. Each link $\{x,\mu\}^*$ of the dual lattice corresponds to the 3-cube $\cC_\mu(x)$ of the original lattice.
Then Eq.~\eq{eq:magnetic} in continuum has the following lattice analogue~\cite{ref:ITEP:M}:
\beqn
\langle S_M\rangle_{\mathrm{mon}} & = & \left\langle\sum\limits_x \sum\limits_{\mu=1}^4 j_\mu^2(x)\right\rangle^{-1}
\label{eq:SM:lat}\\
& & \hskip -15mm
\cdot \left\langle
\sum_x \sum_{\mu=1}^4 j_\mu^2(x) \left[\frac{1}{6} \sum_{P \in \partial \,\cC_\mu(x)} \left(1 - \frac{1}{2} \Tr U_P\right)\right]
\right\rangle\,,
\nonumber
\eeqn
where the sum over plaquettes is taken over all six faces $P$ of the 3-cube $\cC_\mu(x)$.

The lattice definition~\eq{eq:SM:lat} corresponds to a minimal splitting between the points $x$ and $x'$
in the continuum expression~\eq{eq:magnetic}.
The center of the monopole coincides with the center of the 3-cube $\cC_\mu(x)$ which surrounds the monopole.
On the other hand the monopole contribution~\eq{eq:SM:lat} to the magnetic part gluon action density is
defined by the average of the plaquettes at the faces of the cube. Thus, there is a minimal distance
between the monopole position and the point where the gluonic action density is measured. Thus, the lattice
formulation~\eq{eq:SM:lat} of the magnetic correlator~\eq{eq:magnetic} provides us with a natural definition
of the splitting between the points $x$ and $x'$. The minimal distance equals to the half lattice spacing~$a$:
\beqn
r_M(a) = \frac{a}{2}\,.
\label{eq:RM}
\eeqn

The lattice analogue of the chromoelectric correlator~\eq{eq:electric} is~\cite{ref:ITEP:M}:
\beqn
\langle S_E\rangle_{\mathrm{mon}} & = & \left\langle\sum\limits_x \sum\limits_{\mu=1}^4 j_\mu^2(x)\right\rangle^{-1}
\label{eq:SE:lat}\\
& & \hskip -20mm
\left\langle \sum_x \sum_{\mu=1}^4 j_\mu^2(x) \left[\frac{1}{24}
\sum_{P \in \partial \,{\cal P}[\cC_\mu(x)]} \left(1 - \frac{1}{2} \Tr U_P\right)\right]
\right\rangle\,,
\nonumber
\eeqn
where one of the sums is taken over 24 plaquettes $P$ which form a manifold ${\cal P}[\cC_\mu(x)]$ corresponding to the 3-cube $\cC_\mu(x)$.
This manifold is represented by a set of all plaquettes $P$ which satisfy the following two conditions:
\begin{itemize}
\item[(i)] each plaquette have one, and only one, common link $l_\nu$ with the cube $\cC_\mu(x)$;
\item[(ii)] each plaquette is lying in the planes, defined by the vectors $\mu$ (the direction of the monopole current)
and $\nu$ (the direction of the link defined in the previous condition).
\end{itemize}

One can estimate a minimal distance between the gluonic operator $S_P$ and the position of the monopole
in the lattice formulation~\eq{eq:SE:lat} of the electric correlator~\eq{eq:electric}:
\beqn
r_E(a) = \frac{a}{\sqrt{2}}\,.
\label{eq:RE}
\eeqn
In order to make this estimation we assumed that the monopole trajectory is static. This assumption is justified because
as the temperature increases the zeroth Matsubara component for all bosonic fields -- including the monopole trajectories
-- becomes dominant.

The excess of the magnetic and electric parts of the gluonic action due to the presence of the monopoles is given
by the following formulae, respectively,
\beqn
\delta S_A = \left\langle S_A \right\rangle_{\mathrm{mon}} - \left\langle S_A \right\rangle_{\mathrm{vac}}\,,
\qquad A = M,E\,.
\label{eq:excess}
\eeqn
where  $\left\langle S_A \right\rangle_{\mathrm{vac}}$ is the magnetic (for $A=M$) and electric (for $A=E$)
action densities in the vacuum:
\beqn
\left\langle S_M \right\rangle_{\mathrm{vac}} & = & \frac{1}{3 L_s^3 L_t}
\left\langle\sum_{P_{\mathrm{spat}}} \left(1 - \frac{1}{2} \Tr U_{P_{\mathrm{spat}}}\right) \right\rangle\,,
\\
\left\langle S_E \right\rangle_{\mathrm{vac}} & = & \frac{1}{3 L_s^3 L_t}
\left\langle\sum_{P_{\mathrm{temp}}} \left(1 - \frac{1}{2} \Tr U_{P_{\mathrm{temp}}}\right) \right\rangle\,.
\eeqn
Here the sums go over spatial ($P_{\mathrm{spat}}$) and temporal ($P_{\mathrm{temp}}$) plaquettes, respectively.

\begin{figure}[ht]
\begin{center}
  \includegraphics[width=75mm, angle=0]{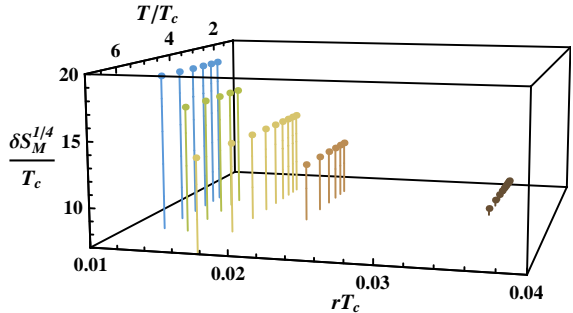}
\end{center}
\vskip -7mm
  \caption{The excess of the magnetic component $\delta S_M$ of the gluonic action~\eq{eq:excess}
   as the function of the distance $r$, Eq.~\eq{eq:RM} from the monopole and the
   temperature. All quantities are given in units of the critical temperature $T_c$.}
  \label{fig:m}
\end{figure}

\vskip 5mm
\subsection{Observation of electromagnetic duality}

In Fig.~\ref{fig:m} we plot the excess of the magnetic component $\delta S_M$
of the gluonic action~\eq{eq:excess}
over the corresponding vacuum expectation value: results are reported in dimensionless
units and as a function of the distance $r$ from the monopole center and of the temperature.
{ We include into this figure all available data sets which are described in Appendix~A
in more details.}
Notice that available data refer to a single value of $r$ for each fixed value of the UV cutoff,
so that results reported in Fig.~\ref{fig:m} have been obtained by collecting together
data from different simulations performed at different values of the lattice spacing. Therefore
the dependence of  $\delta S_M$ on $r$, as reported in Fig.~\ref{fig:m}, cannot be disentangled
from a possible dependence of $\delta S_M$ on the UV cutoff: more simulations in which the
non-abelian action density is measured at more than one value of $r$ for each lattice spacing
would be necessary to completely address this issue. For that reason, our present discussion
about the dependence of $\delta S_M$ on $r$ should be considered as preliminary and
qualitative.

\begin{figure}[ht]
\begin{center}
  \includegraphics[width=75mm, angle=0]{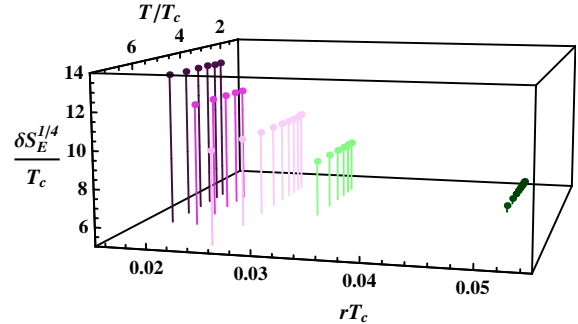}
\end{center}
\vskip -7mm
  \caption{The same as in Fig.~\ref{fig:m} but for the electric part~$\delta S_E$. The distance
  is defined by Eq.~\eq{eq:RE}.}
  \label{fig:e}
\end{figure}

The excess of the magnetic action near thermal monopoles
turns out to be a positive quantity at all distances $r$ from the monopole~\eq{eq:RM}
and at all temperatures, moreover it decreases as a function of $r$. This is an unexpected
result because at zero temperature the condensed monopoles
have an excess density of magnetic action which, even if still positive, increases as a function
of $r$, i.e. it is minimum at the monopole center~\cite{ref:anatomy}. As the monopoles from the condensate evaporate
into the thermal monopoles, one could expect that in the deconfinement phase the excess of the action near monopoles
has the same qualitative features as below $T_c$. This expectation turns out to be not the case, however.
The dependence on $T$ is much milder than that on $r$: the excess of the magnetic action is almost temperature-independent.
The same qualitative features are also seen for the electric part of the action, which we show in Fig.~\ref{fig:e}.

It is interesting to note that the electric and magnetic contributions practically coincide with each other,
\beqn
\langle E^2 \rangle \approx \langle B^2 \rangle\,,
\label{eq:dyonic}
\eeqn
where
\beqn
E^2 = \frac{1}{2} G^2_{0i}(x) \quad \mbox{and} \quad B^2 = \frac{1}{4} G^2_{ij}(x)\,,
\label{eq:EB}
\eeqn
are the electric and action densities, respectively [in our units Eq.~\eq{eq:dyonic} means $\delta S_M \approx \delta S_E$].
{ On the contrary,} for a purely ``chromoelectric'' object, like quark, one expects
\beqn
\langle E^2 \rangle_{\mathrm{quark}} \neq 0\,, \quad \mbox{and} \quad {\langle B^2\rangle }_{\mathrm{quark}} \sim 0\,,
\eeqn
while for a purely ``chromomagnetic'' object, like a monopole, one should get an opposite relation,
\beqn
\langle E^2 \rangle_{\mathrm{mon}} \sim 0\,, \quad \mbox{and} \quad {\langle B^2 \rangle}_{\mathrm{mon}} \neq  0\,.
\eeqn

\begin{figure}[ht]
\begin{center}
  \includegraphics[width=85mm, angle=0]{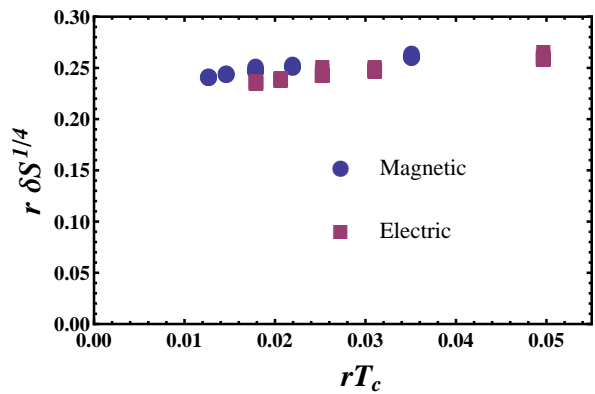}
\end{center}
\vskip -7mm
  \caption{Illustration of the dyonic nature of the objects: The magnetic (the filled circles) and electric (the filled squares)
  contributions to (the square root of) the action density multiplied by the distance $r$ as the function of $r$.
  The statistical error bars are much smaller than the symbols used.}
  \label{fig:c}
\end{figure}

In Fig.~\ref{fig:c} we graphically demonstrate the validity of Eq.~\eq{eq:dyonic} by plotting the overlap
of the Figs.~\ref{fig:m} and \ref{fig:e}. For convenience, we show the both contributions to the energy density
multiplied by the distance $r$. Basically, Fig.~\ref{fig:c} represents a scaled projection of Figs.~\ref{fig:m} and \ref{fig:e}
onto the distance-action plane. This figure illustrates the amazing fact that the electric and magnetic actions are the same around
the ``monopoles''. Indeed, the fact, that the electric and magnetic contributions are almost the same
is unusual because from the first principles we would expect that the magnetic component of the gluonic
action near the monopole is much stronger than the electric part of the gluonic action.

Even if the equality of electric and magnetic contribution is observed only
on average quantities, it is tempting to make the hypothesis that thermal abelian monopoles
are associated to non-abelian field configurations possessing EM duality, i.e. selfdual objects.
Numerically, the deviations from the EM duality do not exceed~3.5\%.

The objects which carry equal amounts of electric and magnetic components are non-Abelian dyons. The (classical) fields
of dyons satisfy the selfduality relation
\beqn
G_{\mu\nu} = {\widetilde G}_{\mu\nu}\,,\qquad {\widetilde G}_{\mu\nu} = \frac{1}{2} \epsilon_{\mu\nu\alpha\beta} G_{\alpha\beta}\,,
\label{eq:G}
\eeqn
so that electric and magnetic action densities~\eq{eq:EB} of the dyons are the same,
\beqn
E^2_{\mathrm{dyon}} = B^2_{\mathrm{dyon}}\,.
\eeqn
This equation is in line with our numerical observation~\eq{eq:dyonic}.

The simplest selfdual configuration is a (static) Bogomolny--Prasad--Sommerfield (BPS) solution
to the classical equations of motion of $SU(2)$ Yang--Mills theory~\cite{ref:BPS}:
\beqn
A^a_i & = & \frac{1}{g}\, f(r) \varepsilon_{iab} \frac{x^b}{r}\,,\quad f(r) = \frac{1}{r} \Bigl(1 - \frac{r}{\sinh r}\Bigr)\,,
\label{eq:BPS:spatial}\\
A^a_4 & = & \frac{1}{g}\, h(r) \, \frac{x^a}{r}\,,\quad \hspace{5mm} h(r) = \frac{1}{r} \Bigl(r \coth r -1 \Bigr)\,.
\label{eq:BPS:temporal}
\eeqn
Here the distance coordinates $r \equiv |{\vec x}|$ and $x$ are scaled by an arbitrary factor, $r \to r_0 \cdot r$,
to make a dimensionless quantity.

\begin{figure}[ht]
\begin{center}
  \includegraphics[width=80mm, angle=0]{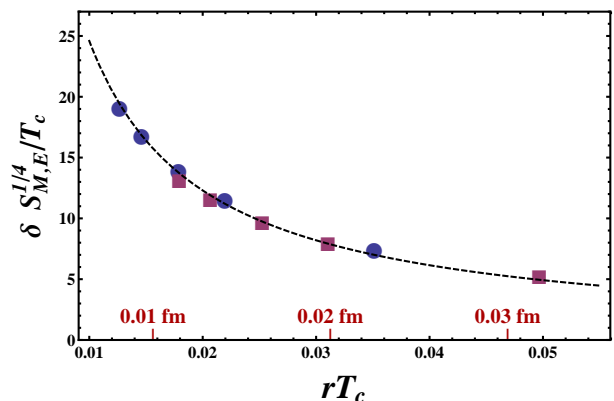}
\end{center}
\vskip -7mm
  \caption{The fit of the excesses in the magnetic (the blue circles) and electric (the red squares)
  action densities near the monopole as the function of the distance $r$ from the monopole (shown both
  in units of $T_c^{-1}$ and in Fermi). The error bars and the magnitude of the temperature dependence
  is much smaller than the size of the symbols. The data are fitted by the function~\eq{eq:r4}. The fit is shown by the dashed line.}
  \label{fig:me:fit}
\end{figure}

In order to characterize both the singularity at the origin of the thermal monopole (dyon) and the observed EM duality
we fit the magnetic and electric excesses of the action density by the function
\beqn
\delta S_\ell = \left(\frac{C_\ell}{r}\right)^4\,.
\label{eq:r4}
\eeqn
We performed three different fits. We fitted the electric action density, $\ell = E$ and, separately, the magnetic action density, $\ell = M$,
and finally, we made the fit of the both { simultaneously}, $\ell=EM$. In all these fits we have averaged our data points corresponding
to different temperatures $T$ and the same distances $r$ (this is a legitimate operation since the data are almost $T$--independent
as one can see from Fig.~\ref{fig:c}). The best fit parameters are, respectively:
\beqn
C_E & = & 0.243(2)\,,
\nonumber\\
C_M & = & 0.247(2)\,,
\\
C_{EM} & = & 0.246(2)\,.
\nonumber
\eeqn
These values are very close to each other, so that the duality~\eq{eq:dyonic} works with a high precision. In Fig.~\ref{fig:me:fit}
we plot the joint fit $\ell=EM$ in order to illustrate the quality of { the selfduality} relation~\eq{eq:dyonic}.
Thus, in this paper we refer to these objects as both monopoles and selfdual dyons simultaneously, because
they are identified as monopoles by their magnetic charge, while their non-Abelian content shows signatures
of the dyonic selfduality.

Note that (smooth) dyonic configurations were found in
Ref.~\cite{ref:dyonic:lattice} using a method of cooling of the
gluon configurations and/or utilizing the low-lying fermion modes to
filter the positions of the dyons. In this paper, assuming that
thermal monopoles are indeed associated with dyons, we report the
{\it direct} observation of dyons in dynamical (uncooled)
configurations. In fact, the field of the classical BPS
dyon~\eq{eq:BPS:spatial}, \eq{eq:BPS:temporal} is a smooth function
at the origin, $r=0$. On the contrary, our data suggest that the
action of the thermal dyons behave as singular as $1/r^4$. One would
conclude that quantum thermal dyons are not resembling the smooth
classical BPS solutions at all. On the other hand, as already
stressed above, our measurements have been taken at the scale of the
lattice spacing, so that the dependence on $r$ cannot really be
disentangled from that on the UV cutoff and our discussion about the
$1/r^4$ behavior should be regarded as only qualitative: more
refined measurements should be performed in order to clarify the
issue.

\subsection{Contribution of thermal monopoles to pressure}

Finally, let us discuss the contribution of thermal monopoles { (or, dyons)} to the thermodynamics of the plasma.
The positive value of the excess of both electric and magnetic action densities near the monopoles
means that their contributions to the
trace anomaly $\theta$, Eqs.~\eq{eq:theta:continuum}, \eq{eq:anomaly:lattice} and, consequently, to the pressure $p$
and to the energy density $\varepsilon$, Eq.~\eq{eq:pressure:anomaly}, are negative quantities~\cite{footnote1}:

\beqn
\delta\theta_{\mathrm{mon}}(T) < 0\,,
\nonumber\\
\delta p_{\mathrm{mon}}(T) < 0\,,
\label{eq:dyon:effect}\\
\delta\varepsilon_{\mathrm{mon}}(T) < 0\,,
\nonumber
\eeqn
because the $\beta$-function $\tilde\beta(g)$ in Eq.~\eq{eq:theta:continuum} is a negative-valued function.
We cannot calculate numerical values of the monopole contribution to these thermodynamic quantities because we should also take into
account the entropy of the monopoles. This task cannot be completed using our method. We can only state the qualitative features of the
effect of the monopoles~\eq{eq:dyon:effect}.

\section{Conclusion}

In this paper we have studied the correlation of thermal abelian monopoles detected after MAG projection
with the gauge invariant nonabelian action. The study has been performed for QCD with two colors and in the quenched limit.
We have found an equal excess of electric and magnetic action around thermal monopoles, suggesting that the underlying
non-abelian gauge configurations -- associated to thermal monopoles -- may correspond to selfdual dyons.
The electric-magnetic equality holds with a good numerical precision (less than 4\%).

We notice that the presence of the selfdual dyons {in the thermal plasma} is a nontrivial fact. Indeed, one can
argue~\cite{ref:dyons:lattice}
that the thermal medium is populated with gluonic monopoles, which become dyons via the Witten effect~\cite{ref:Witten}
as these monopoles pass through self-dual regions (say, instantons) in space-time. In this case the monopole would get a
space-dependent electric charge which, however, will be vanishing far from the instantons. In this scenario the
electric action density should be small compared to the rival magnetic component. On the contrary, we have observed
exact selfduality indicating that -- at least at short distances -- the electric component is as strong
as the magnetic one.
Note that the electric-magnetic duality may be a property peculiar to thermal monopoles observed in the deconfined phase of Yang-Mills theories.

We have found that thermal monopoles (dyons) make a negative contribution to the pressure and to the energy density of the gluon plasma.

Most importantly, certain signatures of the dyons may be observed experimentally in the heavy-ion collisions via the so-called chiral magnetic effect.
Indeed, since the dyons are characterized by nonzero topological charge density, they are playing a role of ``seeds'' of
topological charge fluctuations in the plasma. In a strong magnetic field the topological structures in the gluonic
vacuum leave their footprints via certain $CP$-odd correlations like, for example,
the generation of the electric current due to the chiral magnetic effect~\cite{Kharzeev:08:1}.
Preliminary signatures of such a current were already observed experimentally
by the STAR collaboration at RHIC~\cite{Voloshin:08:1} (this effect was also observed in the lattice
gauge theory in Ref.~\cite{ref:CME:Lattice}).

In order to clarify further the nature of these objects we are planning to calculate
correlations of the thermal monopoles with the local topological charge density. We are also going to
measure cross-correlations of fluctuations of electric and magnetic action densities around the thermal monopoles,
and extend our study of correlation functions to larger distances from the monopoles.

\acknowledgments

This work has been started during the
workshop ``Non-Perturbative Methods in Strongly Coupled Gauge
Theories'' at the Galileo Galilei Institute (GGI) in Florence and
was supported by the STINT Institutional grant IG2004-2 025,
by the RFBR 08-02-00661-a, DFG-RFBR 436 RUS, by a grant for scientific
schools NSh-679.2008.2, by the Federal Program of
the Russian Ministry of Industry, Science and Technology
No. 40.052.1.1.1112 and by the Russian Federal Agency
for Nuclear Power. Numerical simulations have been performed
on a computer farm in Genova provided by INFN.

\appendix
\section{Numerical simulations}
\label{sec:numerical}

Our numerical simulations were performed in $SU(2)$ lattice gauge theory
with the standard plaquette action. The basic setup is the same as
the one adopted in Ref.~\cite{ref:evidence}. We used various lattice geometries
$L_s^3 \times L_t$ with varying temporal extension of the lattice $L_t = 4,\dots 13$,
and fixed spatial extension $L_s = 64$. We performed simulations at different values
of the lattice gauge  coupling~\eq{eq:beta}, $\beta = 2.7,\, 2.86,\, 2.93,\, 3.00,\, 3.05$.

The physical value of the lattice spacing was determined from the relation
\beqn
a(\beta) \Lambda_L = R(\beta) \lambda(\beta)\,,
\eeqn
where $R$ is the two-loop perturbative $\beta$-function,
while $\lambda$ is a
non-perturbative correction factor computed and reported in Ref.~\cite{karsch}. The mass scale $\Lambda_L$
is related to the critical temperature~\cite{karsch}, $T_c = 21.45(14) \Lambda_L$,
and to the zero--temperature string tension $\sigma$ via $T_c / \sqrt{\sigma} = 0.69(2)$~\cite{karsch2}.
The set of our values of $\beta$ corresponds to five lattice spacings in the interval $a = (0.015 \dots 0.045)\,\mbox{fm}$.
The set of temporal extensions $L_t$ and the lattice spacings $a$ determines a grid of temperature values $T = 1/(L_t a)$
which is rather wide, $T/T_c = 1.095, \, \cdots ,\, 5.637$.

The currents of the Abelian monopole were identified by
the standard De Grand-Toussaint construction~\cite{degrand}
after performing an Abelian projection in the Maximal Abelian gauge.
This gauge is defined by the condition of the maximization of the following functional
with respect to gauge transformations:
\begin{equation}
F_{\rm MAG} = \sum_{\mu,x} {\rm Re}\,  \mbox{tr} \left[U_\mu(x)
  \sigma_3 U^{\dagger}_\mu(x) \,
\sigma_3\right]
\label{maxfun}
\end{equation}
The maximization is achieved by an iterative combination of local
maximization and overrelaxation~\cite{cosmai}.

In the Maximal Abelian gauge the monopole currents $j_\mu$ are defined as
location of sources of the magnetic fields in the diagonal components of the gauge fields~\cite{degrand}:
\begin{equation}
j_\mu = {1 \over 2 \pi} \varepsilon_{\mu\nu\rho\sigma} \hat\partial_\nu \overline
\theta_{\rho\sigma}\,.
\end{equation}
Here $\overline \theta_{\mu\nu}$ is a compactified plaquette angle, $ - \pi < \overline \theta_{\mu\nu} \leqslant \pi$,
defined as follows:
\begin{equation}
\theta_{\mu\nu}= \overline \theta_{\mu\nu} + 2 \pi n_{\mu\nu}\,, \qquad n_{\mu\nu}\in \mathbb{N}\,.
\end{equation}


\begin{thebibliography}{99}

\bibitem{review}
D.~E.~Kharzeev,
{``Hot and dense matter: from RHIC to LHC: Theoretical overview''},
  arXiv:0902.2749;\\
E.~Shuryak,
  Prog.\ Part.\ Nucl.\ Phys.\  {\bf 62}, 48 (2009)
  [arXiv:0807.3033].

\bibitem{ref:magn:comp}
M.~N.~Chernodub and V.~I.~Zakharov,
Phys.\ Rev.\ Lett.\  {\bf 98}, 082002 (2007)
[arXiv:hep-ph/0611228].

\bibitem{shuryak}
J.~Liao and E.~Shuryak,
Phys.\ Rev.\  C {\bf 75}, 054907 (2007)
[arXiv:hep-ph/0611131].

\bibitem{ref:evidence}
A.~D'Alessandro, M.~D'Elia,
Nucl.\ Phys.\  B {\bf 799}, 241 (2008) [arXiv:0711.1266] and
arXiv:0812.1867.

\bibitem{EOS:vortices}
M.~N.~Chernodub, A.~Nakamura and V.~I.~Zakharov,
Phys.\ Rev.\  D {\bf 78}, 074021 (2008) [arXiv:0807.5012].

\bibitem{EOS:magnetic}
M.~N.~Chernodub, K.~Ishiguro, A.~Nakamura, T.~Sekido, T.~Suzuki and V.~I.~Zakharov,
PoS {\bf LAT2007}, 174 (2007) [arXiv:0710.2547].

\bibitem{sasha}
A.~Gorsky and V.~Zakharov,
Phys.\ Rev.\  D {\bf 77}, 045017 (2008) [arXiv:0707.1284];
 A.~S.~Gorsky, V.~I.~Zakharov and A.~R.~Zhitnitsky,
  Phys.\ Rev.\  D {\bf 79}, 106003 (2009)
  [arXiv:0902.1842 [hep-ph]].


\bibitem{ref:magn:strings}
M.~N.~Chernodub and V.~I.~Zakharov,
{``Monopoles and vortices in Yang-Mills plasma''},
 arXiv:0806.2874;\\
{``Magnetic strings as part of Yang-Mills plasma''},
[arXiv:hep-ph/0702245].

\bibitem{ref:viscosity}
C.~Ratti and E.~Shuryak,
  Phys.\ Rev.\  D {\bf 80}, 034004 (2009)
  [arXiv:0811.4174 [hep-ph]];
M.~N.~Chernodub, H.~Verschelde and V.~I.~Zakharov,
  arXiv:0905.2520 [hep-ph];

\bibitem{ref:dual}
G.~'t~Hooft, in {\it High Energy Physics}, ed. A. Zichichi,
EPS International Conference, Palermo (1975);\\
S.~Mandelstam, {Phys.\ Rept.}  {\bf 23}, 245 (1976).

\bibitem{superI-II}
A.~Di Giacomo, B.~Lucini, L.~Montesi and G.~Paffuti,
  Phys.\ Rev.\  D {\bf 61}, 034503 (2000)
  [arXiv:hep-lat/9906024];
  Phys.\ Rev.\  D {\bf 61}, 034504 (2000)
  [arXiv:hep-lat/9906025].

\bibitem{superIV}
  M.~D'Elia, A.~Di Giacomo, B.~Lucini, G.~Paffuti and C.~Pica,
  Phys.\ Rev.\  D {\bf 71}, 114502 (2005)
  [arXiv:hep-lat/0503035].

\bibitem{moscow}
  M.~N.~Chernodub, M.~I.~Polikarpov and A.~I.~Veselov,
  Phys.\ Lett.\  B {\bf 399}, 267 (1997)
  [arXiv:hep-lat/9610007].


\bibitem{bari}
  P.~Cea and L.~Cosmai,
  JHEP {\bf 0111}, 064 (2001);
  P.~Cea, L.~Cosmai and M.~D'Elia,
  JHEP {\bf 0402}, 018 (2004)
  [arXiv:hep-lat/0401020].

\bibitem{greensite}
  J.~Greensite,
  Prog.\ Part.\ Nucl.\ Phys.\  {\bf 51}, 1 (2003)
  [arXiv:hep-lat/0301023].

\bibitem{ref:preliminary}
V.~G.~Bornyakov, V.~K.~Mitrjushkin, M.~Muller-Preussker,
Phys.\ Lett.\  B {\bf 284}, 99 (1992);
S.~Ejiri,
Phys.\ Lett.\  B {\bf 376}, 163 (1996) [arXiv:hep-lat/9510027].

\bibitem{footnote:dyon}
{ To be more precise, in this paper we use the word ``dyon'' in order to describe an object in the Euclidean spacetime,
which is characterized by the same (on the quantitative level) $1/r^{-4}$ behavior of the electric and magnetic
nonabelian action densities.}

\bibitem{ref:dyons:coninuum}
  A.~Gonzalez-Arroyo and Yu.~A.~Simonov,
  Nucl.\ Phys.\  B {\bf 460}, 429 (1996)
  [arXiv:hep-th/9506032].

\bibitem{ref:dyons:lattice}
 V.~Bornyakov and G.~Schierholz,
  Phys.\ Lett.\  B {\bf 384}, 190 (1996)
  [arXiv:hep-lat/9605019];
  M.~N.~Chernodub, F.~V.~Gubarev and M.~I.~Polikarpov,
  JETP Lett.\  {\bf 69}, 169 (1999)
  [arXiv:hep-lat/9801010].

\bibitem{ref:dyonic:lattice}
  V.~G.~Bornyakov, E.~M.~Ilgenfritz, B.~V.~Martemyanov and M.~Muller-Preussker,
  Phys.\ Rev.\  D {\bf 79}, 034506 (2009)
  [arXiv:0809.2142 [hep-lat]];
 V.~G.~Bornyakov, E.~M.~Ilgenfritz, B.~V.~Martemyanov, S.~M.~Morozov, M.~Muller-Preussker and A.~I.~Veselov,
  Phys.\ Rev.\  D {\bf 76}, 054505 (2007)
  [arXiv:0706.4206 [hep-lat]];
 E.~M.~Ilgenfritz, B.~V.~Martemyanov, M.~Muller-Preussker and A.~I.~Veselov,
  Phys.\ Atom.\ Nucl.\  {\bf 68}, 870 (2005).

\bibitem{Kharzeev:08:1}
  D.E.~Kharzeev, L.D.~McLerran, and H.J.~Warringa,
  Nucl.\ Phys.\  A {\bf 803}, 227 (2008) [arXiv:0711.0950];
 K.~Fukushima, D.E.~Kharzeev, and H.J.~Warringa,
  Phys.\ Rev.\  D {\bf 78}, 074033 (2008) [arXiv:0808.3382].

\bibitem{Voloshin:08:1}
  S.~A.~Voloshin  [STAR Collaboration],
  arXiv:0806.0029 [nucl-ex];
H.~Caines [STAR Collaboration], arXiv:0906.0305 [nucl-ex].

\bibitem{ref:CME:Lattice}
P.~V.~Buividovich, M.~N.~Chernodub, E.~V.~Luschevskaya and M.~I.~Polikarpov,
  Phys. Rev. D {\bf 80}, 054503 (2009)
  [arXiv:0907.0494]; Pis'ma v ZhETF {\bf 90}, 456 (2009).

\bibitem{Yee:2009vw}
  H.~U.~Yee,
  ``Holographic Chiral Magnetic Conductivity,''
  arXiv:0908.4189 [hep-th];
  A.~Rebhan, A.~Schmitt and S.~A.~Stricker,
  arXiv:0909.4782 [hep-th].

\bibitem{Boyd:1996bx}
  G.~Boyd, J.~Engels, F.~Karsch, E.~Laermann, C.~Legeland, M.~Lutgemeier and B.~Petersson,
  Nucl.\ Phys.\  B {\bf 469}, 419 (1996)
  [arXiv:hep-lat/9602007].

\bibitem{ref:bornyakov}
V.~G.~Bornyakov, V.~K.~Mitrjushkin and M.~Muller-Preussker,
  Phys.\ Lett.\  B {\bf 284}, 99 (1992).

\bibitem{ref:ejiri}
S.~Ejiri,
  Phys.\ Lett.\  B {\bf 376}, 163 (1996)
  [arXiv:hep-lat/9510027].

\bibitem{ref:ITEP:M}
B.~L.~G.~Bakker, M.~N.~Chernodub and M.~I.~Polikarpov,
Phys.\ Rev.\ Lett.\  {\bf 80}, 30 (1998)
[arXiv:hep-lat/9706007].

\bibitem{ref:ITEP:E}
B.~L.~G.~Bakker, M.~N.~Chernodub, M.~I.~Polikarpov, A.~I.~Veselov,
Phys.\ Lett.\  B {\bf 449}, 267 (1999) [arXiv:hep-lat/9811001].

\bibitem{ref:anatomy}
V.~G.~Bornyakov, M.~N.~Chernodub, F.~V.~Gubarev, M.~I.~Polikarpov, T.~Suzuki, A.~I.~Veselov and V.~I.~Zakharov,
  Phys.\ Lett.\  B {\bf 537}, 291 (2002)
  [arXiv:hep-lat/0103032].

\bibitem{ref:BPS}
E.~B.~Bogomolny, Sov.\ J.\ Nucl.\ Phys.\  {\bf 24}, 449 (1976);
M.~K.~Prasad and C.~M.~Sommerfield,
Phys.\ Rev.\ Lett.\  {\bf 35}, 760 (1975);
P.~Rossi, Phys.\ Rept.\  {\bf 86}, 317 (1982).

\bibitem{footnote1}{The positive value
of the monopole contribution reported in Ref.~\cite{EOS:magnetic} can be explained by the fact that in that study the
contribution of the ultraviolet monopoles was not subtracted.}

\bibitem{ref:Witten}
  E.~Witten,
  Phys.\ Lett.\  B {\bf 86}, 283 (1979).

\bibitem{karsch}
J.~Engels, F.~Karsch and K.~Redlich,
  Nucl.\ Phys.\  B {\bf 435}, 295 (1995)
  [arXiv:hep-lat/9408009].

\bibitem{karsch2}
J.~Fingberg, U.~M.~Heller and F.~Karsch,
  Nucl.\ Phys.\  B {\bf 392}, 493 (1993)
  [arXiv:hep-lat/9208012].

\bibitem{degrand}
T.~A.~DeGrand and D.~Toussaint,
  Phys.\ Rev.\  D {\bf 22}, 2478 (1980).

\bibitem{cosmai}
P.~Cea and L.~Cosmai,
  Phys.\ Rev.\  D {\bf 52}, 5152 (1995)
  [arXiv:hep-lat/9504008].

\end{thebibliography}
\end{document}